\documentclass[twocolumn,showkeys,preprintnumbers,amsmath,amssymb,prb]{revtex4}

\usepackage[dvips]{graphicx}
\usepackage{dcolumn}
\usepackage{bm}
\usepackage{subfigure}

\begin{document}

\preprint{published in The Journal of Chemical Physics, J. Chem. Phys. {\bf
    122}, p. 054501 (2005), doi:10.1063/1.1839574 }

\title{Breakdown of the mirror image symmetry in the optical absorption/emission spectra of oligo(\textit{para}-phenylene)s}

\author{Georg Heimel}
\email{georg.heimel@chemistry.gatech.edu} \affiliation{School of
Chemistry and Biochemistry, Georgia Institute of Technology,
Atlanta, Georgia 30332-0400\\ Institute of Solid State Physics,
Graz University of
Technology, Petersgasse 16, A-8010 Graz, Austria}%

\author{Maria Daghofer}
\affiliation{Institute of Theoretical and Computational Physics,
Graz University of Technology, Petersgasse 16, A-8010 Graz,
Austria}

\author{Johannes Gierschner}
\affiliation{Institute of Physical and Theoretical Chemistry,
University of T\"ubingen, Auf der Morgenstelle 8, D-72076
T\"ubingen, Germany}

\author{Emil J. W. List}
\affiliation{Christian Doppler Laboratory Advanced Functional Materials\\
Institute of Solid State Physics, Graz University of Technology,
Petersgasse 16, A-8010 Graz, Austria\\ Institute of Nanostructured
Materials and Photonics, Joanneum Research, A-8160 Weiz}

\author{Andrew C. Grimsdale}
\author{Klaus M\"ullen}
\affiliation{Max-Planck-Institut f\"ur Polymerforschung, Ackermannweg 10, D-55128 Mainz, Germany}

\author{David Beljonne}
\affiliation{Service de Chimie des Mat\'eriaux Nouveaux, Centre de
Recherche en Electronique et Photonique Mol\'eculaires,
Universit\'e de Mons-Hainaut, B-7000 Mons, Belgium}

\author{Jean-Luc Br\'edas}
\affiliation{School of Chemistry and Biochemistry, Georgia Institute of Technology, Atlanta, Georgia 30332-0400}%

\author{Egbert Zojer}
\affiliation{Institute of Solid State Physics, Graz University of
Technology,
Petersgasse 16, A-8010 Graz, Austria\\
School of Chemistry and Biochemistry, Georgia Institute of
Technology, Atlanta, Georgia 30332-0400}%

\date{\today}

\begin{abstract}
The absorption and emission spectra of most luminescent,
$\pi$-conjugated, organic molecules are the mirror image of each
other. In some cases, however, this symmetry is severely broken.
In the present work, the asymmetry between the absorption and
fluorescence spectra in molecular systems consisting of
\textit{para}-linked phenyl rings is studied. The vibronic
structure of the emission and absorption bands is calculated from
\textit{ab-initio} quantum chemical methods and a subsequent,
rigorous Franck-Condon treatment. Good agreement with experiment
is achieved. A clear relation can be established between the
strongly anharmonic double-well potential for the phenylene ring
librations around the long molecular axis and the observed
deviation from the mirror image symmetry. Consequences for related
compounds and temperature dependent optical measurements are also
discussed.

\end{abstract}

\maketitle

\section{\label{Intro}Introduction}
The poly(\textit{para}-phenylene)s and their planarized
derivatives are of particular interest in the field of
\mbox{$\pi$-conjugated} materials due to their use in
opto-electronic
applications.\cite{LOPPLaserFlex1,PPPLed1,PfoDevice1,PfoDevice2,PfoDevice3,PfoDevice4}
The coupling between electronic and vibrational degrees of freedom
strongly affects the nature of the photoexcitations in these
materials. Thus, a good understanding of this coupling mechanism
is highly desirable to tune the fundamental materials properties.

Because of strong electron-phonon coupling, organic
\mbox{$\pi$-conjugated} molecules are subject to structural
distortions upon photoexcitation. Different Born-Oppenheimer
potential energy surfaces in the ground state (GS) and
electronically excited state (ES) lead to a reorganization of the
molecular geometry in the ES in conjunction with an energetic
stabilization. This strong electron-vibration coupling
dramatically influences the bandshapes of the electronic
transitions in absorption and emission, leading to a pronounced
vibronic structure in the optical spectra. Usually, the major
distortion upon photoexcitation can be attributed to a change in
the C=C bond alternation along the backbone. The related C=C
stretch vibrations lie in the energy range between 1200-1600
cm$^{-1}$ and can be clearly resolved in (resonant) Raman
spectra.\cite{OppModes3,OppModes4,3pRamanLt1,OppModes1,Reli2,LpppRaman1}
The optical spectra exhibit an extended vibronic progression
associated with these modes.

In a simple Franck-Condon (FC) picture, the absorption and
emission spectra are expected to be the mirror images of each
other. Experimentally, one observes mirror imaged absorption and
emission spectra with a sharply resolved vibronic progression in
molecules (e.g. oligoacenes)\cite{OaOptics} and polymers (e.g.
ladder-type
poly(\textit{para}-phenylene))\cite{LpppBands1,LpppBands2,LpppBands3}
that have a rigid backbone. For species that comprise torsional
degrees of freedom in their conjugated backbone, varying degrees
of deviation from the mirror image behavior can be observed. At
room temperature, poly(\textit{para}-phenylenevinylene) (PPV) and
its oligomers exhibit a strong deviation from mirror image
symmetry with the vibronic structure in the absorption spectrum
being rather blurred and smeared out while the vibronic
progression in the emission spectrum is relatively
well-resolved.\cite{PpvBands1,PpvBands2,PpvBands3} At low
temperature, however, the absorption spectrum gradually becomes
more structured and mirror symmetry is
recovered.\cite{Gierschner,Matrix7} Poly(\textit{para}-phenylene)
(PPP) and its oligomers\cite{Reli1,OppOptics} as well as
poly(fluorene)\cite{PfoBands1,PfoBands2,PfoBands3,PfoBands4,EmilPfoReview}
exhibit a broad, completely featureless and unstructured
absorption band while in emission a clearly resolved vibronic
progression appears. It is the purpose of this work to investigate
this strong deviation from the mirror image symmetry for systems
constituted of \textit{para}-linked phenylene rings, addressing
also the influence of temperature.

For PPV-type polymers and PPP, the broadening of the absorption
has been explained by the presence of a distribution of site
energies associated with different conjugation
lengths.\cite{LengthDistrib1,LengthDistrib2,LengthDistrib3,LengthDistrib4,RamanModel5,RamanModel6}
As a consequence, one would observe a superposition of
vibronically structured absorption spectra with different
electronic origins, resulting in a featureless absorption band.
Since exciton migration towards the lowest energy sites (longest
conjugated segments) can take place prior to emission,
fluorescence would then come mainly from a single species (the
longest conjugated segments) with well-defined electronic
transition energy. Thus, in emission, one would observe the
sharply resolved vibronic progression of that species only.

Although this effect is very likely to occur in polymers, it might
not necessarily be the dominant one for PPV and its
oligomers.\cite{FcCalcs1} In this case, the room-temperature
blurring of the absorption spectra can be associated with the
thermal activation of low-frequency librational
modes.\cite{Gierschner} The approach pursued in the present study
indicates that this explanation does not hold for PPP-related
systems. In the latter, the strongly anharmonic potential for the
ring librations can be held responsible for the dramatic breakdown
of the mirror image symmetry in the absorption and emission
spectra, as suggested in Ref. \onlinecite{OppFc2}.

\begin{figure}[!t]
\begin{center}
\subfigure[\textit{p}-terphenyl]{\includegraphics[scale=0.8]{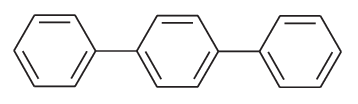}}\\
\subfigure[indenofluorene]{\includegraphics[scale=0.8]{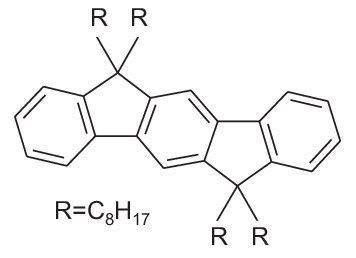}}
\caption{\label{FigMater} Chemical structures of the model
compounds \textit{p}-terphenyl (a) and indenofluorene (b)
investigated in this study.}
\end{center}
\end{figure}

In order to shed light on the influence of the low-energy
torsional vibrations of the phenylene rings on the asymmetry
between absorption and emission spectra, we investigated
\mbox{\textit{p}-terphenyl} (3P) and indenofluorene
(3F)\cite{3LoppSynth} (see Fig. \ref{FigMater}). These systems are
related to the more technologically relevant polymeric species,
ladder-type poly(\textit{para}-phenylene) and
poly(fluorene).\cite{LOPPLaserFlex1,PPPLed1,PfoDevice1,PfoDevice2,PfoDevice3,PfoDevice4}
The electronic structure arising from the conjugated backbone is
nearly identical for the two compounds. The important difference
is that the phenylene rings are able to librate around the long
molecular axis in 3P, while the additional methyl bridges in 3F
keep the backbone rigidly planar and thus effectively block any
torsional motion.\cite{OppFc1} The conjugation length is
well-defined in oligomers, thereby leading to a narrower density
of states with respect to polymers. Consequently, exciton
migration phenomena can be expected to play a minor role for the
shape of the optical spectra in oligomers. In addition, the
experiments reported here have been conducted in dilute solution
where energy transfer is slow compared to radiative decay.

The aim of this study is to show that a rigorous FC treatment of
both high-energy C=C stretching vibrations and low-energy ring
librations in 3P can account for the breakdown of the mirror image
symmetry in the case of 3P as opposed to 3F, where the mirror
symmetry is retained. All required parameters are obtained from
high-level \textit{ab-initio} calculations. The strongly
anharmonic torsional potential in 3P is treated by numerically
evaluating the librational energy levels, wavefunctions, and FC
overlap integrals in curvilinear internal coordinates.
Furthermore, we underline the importance of this procedure by
comparing the results to a strictly harmonic approximation for the
GS torsional potential well.

\section{Theory}
\label{Theory} Within the Born-Oppenheimer and Franck-Condon
approximation, the probability $P_{ge}(E)$ for absorption of a
photon with energy $E$ that promotes the molecule from the GS into
its first ES can be split into an electronic part and a
vibrational part:
\begin{equation}
\label{EqBO} P_{ge}(E)\propto (\mu_{el}^0)^2
P_{\mathbf{v'}\mathbf{v''}}(E)
\end{equation}
where the electric transition dipole matrix element $(\mu_{el}^0)$
is evaluated at the ground-state equilibrium conformation of the
molecule. The oscillator strength of the electronic transition is
redistributed among all the transitions between the vibrational
manifold of the ground and excited state. Assuming that the
vibrational Hamiltonian of a molecule of $N$ nuclei can be
separated into $3N-6$ internal normal coordinates $Q_i$, the total
vibrational wavefunction can be written as
$|\mathbf{v'}\rangle=|v'_1\rangle |v'_2\rangle\ldots
|v'_{3N-6}\rangle$ and $|\mathbf{v''}\rangle=|v''_1\rangle
|v''_2\rangle\ldots |v''_{3N-6}\rangle$ for the ground and excited
state, respectively. The $v_i'$ and $v''_i$ are the vibrational
quantum numbers of the individual oscillators, the total set of
quantum numbers is denoted as $\mathbf{v'}$ and $\mathbf{v''}$.
The vibrational part of the transition probability,
$P_{\mathbf{v'}\mathbf{v''}}$, depends on the incoming photon
energy $E$. Assuming a Gaussian broadening for all individual
transitions, it can be written as:
\begin{eqnarray}
\label{EqFullProb} P_{\mathbf{v'}\mathbf{v''}}(E) & \propto &
\sum_{v_1'}^\infty\ldots\sum_{v_{3N-6}'}^\infty\sum_{v_1''}^\infty\ldots\sum_{v_{3N-6}''}^\infty
B(T)\left |F_{\mathbf{v}'}^{\mathbf{v}''}
 \right |^2\\
 &\times & G\left (E;\Gamma,E_{el}+\sum_{i=1}^{3N-6}\left (\varepsilon_i''(v_i'')-\varepsilon_i'(v_i')
 \right)\right )\nonumber
\end{eqnarray}
$B(T)$ is a Boltzmann weighting factor that determines the
population of the vibrational states within the manifold of the
initial electronic state (GS for absorption, ES for emission) at
finite temperature:
\begin{equation}
\label{EqBoltzmann} B(T)=\left \{\begin{array}{ll}
\frac{1}{Z}exp\left (-\frac{\sum_{i=1}^{3N-6}\varepsilon_i'(v'_i)-\varepsilon_i'(0)}{k_{B}T}\right ) & \text{abs.}\vspace{5mm}\\
\frac{1}{Z}exp\left
(-\frac{\sum_{i=1}^{3N-6}\varepsilon_i''(v''_i)-\varepsilon_i''(0)}{k_{B}T}\right
) & \text{em.}
\end{array}\right .
\end{equation}
where $Z$ is a normalization factor.

The $G$-Term on the far right hand side of Eq. \ref{EqFullProb}
describes a normalized, Gaussian lineshape, which is a function of
$E$, has a width (standard deviation) $\Gamma$, and is centered
around the energy given by the third argument. $E_{el}$ stands for
the energy of the electronic origin (transition energy between the
minima of the ground and excited state parabolas) and
$\varepsilon'_i$ is the energy eigenvalue of the $i^{th}$
oscillator along $Q_i'$ containing $v'_i$ quanta.

The term $F_{\mathbf{v}'}^{\mathbf{v}''}$ in Eq. \ref{EqFullProb}
describes the FC overlap between the total vibronic wavefunctions
of the initial and final states:
\begin{eqnarray}
\label{EqFcGeneral} F_{\mathbf{v}'}^{\mathbf{v}''} & = & \langle
v''_1(Q''_1)\ldots
v''_{3N-6}(Q''_{3N-6})|\nonumber\\
& & \times |v'_1(Q'_1)\ldots v'_{3N-6}(Q'_{3N-6})\rangle
\end{eqnarray}
In the general case, the $Q'_i$ in the GS differ from the set of
$Q''_i$ in the ES. They are related by:
\begin{equation}
\label{EqDuschinsky}
\mathbf{Q}'=\mathbb{D}\mathbf{Q}''+\Delta\mathbf{Q}''
\end{equation}
where $\mathbf{Q}'$ and $\mathbf{Q}''$ abbreviate the complete set
of $3N-6$ internal normal coordinates ($Q'_i$ and $Q''_i$) in the
GS and ES, respectively. $\Delta\mathbf{Q}''$ is the displacement
between the equilibrium geometries in the electronic GS and ES
expressed in the basis of the $Q''_i$ and the rotation between the
GS and ES normal coordinates ($\mathbf{Q}'$ and $\mathbf{Q}''$) is
given by $\mathbb{D}$, the \emph{Duschinsky} matrix. If
$\mathbb{D}\approx\mathbf{1}$ (which will be assumed in the
present study), the multi-center integral in Eq. \ref{EqFcGeneral}
can be factorized as:
\begin{eqnarray}
\label{EqFcFactors1} \tilde{F}_{\mathbf{v}'}^{\mathbf{v}''} & = &
\prod_{i=1}^{3N-6}\langle v''_i(Q_i'')|v'_i(Q''_i+\Delta
Q''_i)\rangle=\nonumber\\
 & = & \prod_{i=1}^{3N-6}\langle v''_i(Q_i''-\Delta
Q''_i)|v'_i(Q''_i)\rangle\
\end{eqnarray}
In order to take partly into account the changes in interatomic
force constants upon photoexcitation, two strategies are pursued:
(i) The first is to project the geometry distortion between the GS
and ES equilibrium geometries, $\Delta\mathbf{Q}$, onto the normal
modes of the accepting electronic state ($\Delta\mathbf{Q}''$ for
absorption and $\Delta\mathbf{Q}'$ for emission). (ii) Given that
the Cartesian atomic displacements $X_i$ are related to the normal
coordinates via $Q_i=\sqrt{\bar{m}_i} X_i$, with $\bar{m}_i$ being
the effective mass of the $i^{th}$ mode, the FC overlap matrix
elements of Eq. \ref{EqFcFactors1} can be rewritten:
\begin{subequations}
\label{EqFcFactors2}
\begin{eqnarray}
\label{EqFcFactors2a} F_{\mathbf{v}'}^{\mathbf{v}''} & = &
\prod_{i=1}^{3N-6}\langle
v''_i(Q_i''-\Delta Q''_i)|v'_i(\sqrt{\bar{m}'_i} X''_i)\rangle\\
\label{EqFcFactors2b}F_{\mathbf{v}'}^{\mathbf{v}''} & = &
\prod_{i=1}^{3N-6}\langle v'_i(Q_i'+\Delta
Q'_i)|v''_i(\sqrt{\bar{m}''_i} X'_i)\rangle
\end{eqnarray}
\end{subequations}
where Eq. \ref{EqFcFactors2a} holds for absorption and Eq.
\ref{EqFcFactors2b} for emission. For parabolic potentials, the
vibronic wavefunctions $|v_i\rangle$ are known (harmonic
oscillator wavefunctions) and the FC factors in Eq.
\ref{EqFcFactors2} can be readily evaluated, allowing for
different frequencies and different effective masses of GS and ES
vibrational states. For general potentials, one needs to resort to
numerical procedures (\textit{vide infra}).

When calculating absorption spectra from Eq. \ref{EqBO}, it needs
to be taken into account that the absorption coefficient
$\alpha(E)$ is related to the transition probability $P_{ge}(E)$
via:
\begin{equation}
\label{EqAbs} \alpha(E)=E P_{ge}(E)
\end{equation}
using the first lines of Eqs. \ref{EqBoltzmann} and
\ref{EqFcFactors2} for the calculation of
$P_{\mathbf{v'}\mathbf{v''}}(E)$ (Eq. \ref{EqFullProb}).

The transition probability for spontaneous emission, $P_{eg}(E)$,
relates to $P_{ge}(E)$ as:\cite{EinsteinCoeff}
\begin{equation}
\label{EqEm1} P_{eg}(E)\propto E^3P_{ge}(E)
\end{equation}
using the second lines of Eqs. \ref{EqBoltzmann} and
\ref{EqFcFactors2} for the calculation of
$P_{\mathbf{v'}\mathbf{v''}}(E)$ (Eq. \ref{EqFullProb}).

Finally, if one detects fluorescence intensity $I_{eg}(E)$ per
energy interval rather than the number of emitted particles, the
measured signal is given by:
\begin{equation}
\label{EqEm2} I_{eg}(E)=E P_{eg}(E) \propto E^4P_{ge}(E)
\end{equation}

\section{\label{Method}Methodology}
\subsection{\label{ExpMeth}Experimental}
Absorption spectra of indenofluorene\cite{3LoppSynth} have been
recorded with a Perkin-Elmer $\lambda$-9 UV/VIS/NIR spectrometer.
Fluorescence spectra have been taken with a Shimadzu RF5301
spectrophotofluorometer that has been calibrated for spectral
sensitivity with an Ocean Optics LS-1-CAL calibration light
source. Measurements have been conducted at ambient temperature in
dilute ($< 1 \mu$g/ml) $CH_2Cl_2$ solution. Room temperature
absorption and fluorescence spectra of \textit{p}-terphenyl have
been taken from Ref. \onlinecite{OppOptics}.

For low temperature fluorescence measurements on
\mbox{\textit{p}-terphenyl}, it was dissolved in \textit{n}-decane
to obtain a solution with absorbance E = 0.05. The sample was
placed in a 0.5 cm cylindrical quartz cell of Suprasil quality in
a closed cycle He cryostat (Kryogenics Technology, model 501A) and
measured on a Spex Fluorolog 222 spectrofluorimeter in
backscattering configuration.

\subsection{\label{CompMeth}Computational}
\subsubsection{Quantum chemical approach}
In order to obtain a reliable description of the geometric
deformation taking place upon exciting the molecules from the GS
into the ES, both electronic states have to be treated at a
comparable level of theory. Therefore, GS energies, equilibrium
geometries, normal coordinates, and vibrational frequencies were
calculated within the Hartree-Fock (HF) approximation. ES
energies, equilibrium geometries, normal coordinates, and
vibrational frequencies were computed by employing a configuration
interaction scheme on top of the HF calculation considering only
singly excited Slater determinants (CIS).\cite{CIS} The core
orbitals were exempt from the variational CI procedure, all
valence and virtual orbitals were included. A 6-311+G(d,p) basis
set \cite{Stern1,6-311Gdp1,6-311Gdp2,plus} was used. For the
subsequent calculation of absorption and emission bandshapes, the
vibrational frequencies were scaled with a common factor of
0.9.\cite{Gaussian} All calculations were performed with
\textsc{Gaussian98} \cite{G98} using tight convergence criteria.
The symmetry of both 3F and 3P was constrained to $C_{2h}$
throughout the calculations, forcing the individual phenylene
rings to be strictly planar.

\subsubsection{Franck-Condon treatment}
For the high-energy, in-plane C=C stretching vibrations of 3F and
3P, harmonic potentials and thus harmonic oscillator-type
wavefunctions were assumed. The mass-weighted interstate
distortion was projected onto the normal modes of the accepting
electronic state in order to determine $\Delta Q_i$ (see Eqs.
\ref{EqDuschinsky} to \ref{EqFcFactors2}). The five most important
modes (largest $\Delta Q_i$) were taken into account, covering
more than 95\% of the total interstate distortion. The first 5
vibrational states ($v_i=0\ldots 4$) of each mode were included in
the calculation, yielding sufficiently converged results. All
transitions with a FC weight of greater than 10$^{-4}$ of the
maximum weight transition have been evaluated. The width of the
Gaussian bandshapes, $\Gamma$ (see Eq. \ref{EqFcGeneral}), was set
to 0.1 eV in both molecules for better comparison with experiment.

\subsubsection{Torsional potentials}
\begin{figure}[!t]
\begin{center}
\includegraphics[scale=0.84]{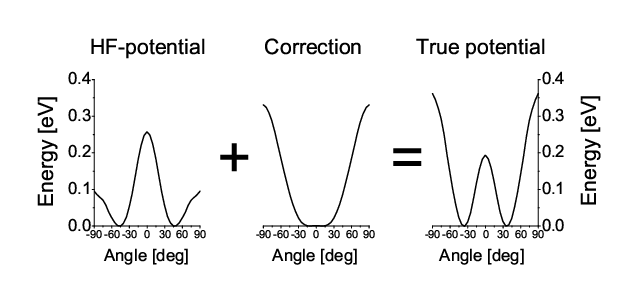}
\caption{\label{FigModDW} Sketch of the process by which (in order
to correct for methodological deficiencies and the influence of
the solvent) a single-well potential (middle panel) has been added
to the Hartree-Fock potential (left panel) in order to give the
actual ground-state torsional potential for \textit{p}-terphenyl
(right panel). The energy scale is the same for all three graphs.}
\end{center}
\end{figure}

As will be shown in the present study, the librational motion of
the rings around the long molecular axis needs to be included for
3P in order to reproduce the experimentally observed bandshapes in
absorption and emisssion. To that end, the diabatic torsional
potentials in the electronic GS and ES of 3P have been computed on
a 5$^\circ$ grid in the interval [-90$^\circ$,90$^\circ$] assuming
alternating signs for the inter-ring tilt angle. For the GS, we
find a $W$-shaped, double-well potential with its minima at $\sim
45^{\circ}$ in accordance with results for biphenyl in gaseous
phase.\cite{2pGasStruct1,2pGasStruct2} However, the experimental
observations we seek to explain in this work have been obtained in
solution, where the inter-ring tilt angle in 3P is experimentally
found to be $\sim 30^{\circ}$ due to the hindrance imposed by the
solvent molecules\cite{3pSolvStruct1,3pSolvStruct2} in the form of
an additional, $U$-shaped potential well. Furthermore, the
rotational barrier at $0^{\circ}$ in solution is lower than that
at $ 90^{\circ}$.\cite{3pSolvStruct1,3pSolvStruct2} In order to
account for the influence of the solvent and the notorious
overestimation of the central rotational barrier in
HF,\cite{TorsionCalcHF} an additional single-well potential needs
to be added to the shallow HF double-well potential (see Fig.
\ref{FigModDW}).
This additional, $U$-shaped potential cannot be derived from
first-principles in a straightforward manner. As a tentative
approximation, we chose to add 20\% of the torsional potential in
the ES to 80\% of the original HF potential well of the GS.

The resulting double-well potential for the GS features a lower
energy barrier at $0^\circ$ tilt angle ($\sim 0.19$ eV) than at
$90^\circ$ ($\sim 0.36$ eV) and the energy minima are pushed
closer together ($\sim 35^{\circ}$) towards the experimentally
determined values.

We would like to point out that the main results of the present
work are not affected by the exact shape of this additional,
\textit{U}-shaped torsional potential. Although quantitative
agreement with experiment (e.g. Stokes shift) slightly depends on
the specific choice for the correcting potential, any GS
double-well deep enough to sustain librational states with an
energy lower than the central rotational barrier leads to the same
qualitative picture for the observed phenomena.

The ES torsional potential is $U$-shaped and almost parabolic. It
exhibits much higher flanks than the GS potential, rendering it
less susceptible to influences from the surroundings. Therefore,
the ES torsional potential has been taken as produced by the CIS
calculations.

\subsubsection{Numerical treatment of the libration}
In order to treat low-energy vibrations with an unusually large
displacement $\Delta Q$ along the internal coordinate \emph{tilt
angle} ($\phi$), one has to resort to curvilinear coordinates as
opposed to linear Cartesian displacements in the case of the
high-energy in-plane C=C strech vibrations of the backbone.
Failing to do so would result in strong but unphysical mode mixing
of the torsion with in-plane stretch motions perpendicular to the
long molecular axis.\cite{OppFc1,Duschin} Furthermore, the GS
torsional double-well potential in 3P is strongly anharmonic. It
has been shown\cite{NumVib1,NumVib2,2pTorsionExpSsJet} that
anharmonic multiple-well potentials can lead to unconventional and
strongly asymmetric vibronic progressions in \emph{soft}
conjugated molecules. Moreover, these authors suggested that a
numerical evaluation of the vibrational wavefunctions and the FC
overlap integrals leads to a better description of the
experimental spectra.

The angular ($\phi$) part of the librational Hamiltonian $H$ can
be written (in cylindrical coordinates and adopting atomic units):
\begin{equation}
\label{EqContHamil}
H=\frac{1}{2I}\frac{\partial^2}{\partial\phi^2}+W(\phi)
\end{equation}
The respective moments of inertia $I$ are extracted from the GS
and ES equilibrium geometries. In order to solve the Schr\"odinger
equation:
\begin{equation}
\label{EqContSchroed} H|v\rangle=\varepsilon |v\rangle
\end{equation}
the angle $\phi$ is discretized:
\begin{eqnarray}
\phi\rightarrow \phi_i & = & i\delta_\phi\nonumber\\
i & = & 1,2,\ldots n\\
\delta_\phi & = & \frac{\phi_{max}}{n}\nonumber
\end{eqnarray}
where $\phi_{max}=90^{\circ}$ in the particular case treated in
the present work and $n=200$ was found to yield fully converged
results. Subsequently, the librational wavefunction $|v\rangle$
and the potential need also to be defined on a discrete grid:
\begin{subequations}
\begin{eqnarray}
v(\phi)\rightarrow v_i & := & v(\phi_i)\\
W(\phi)\rightarrow W_i & := & W(\phi_i)
\end{eqnarray}
\end{subequations}
The derivative in Eq. \ref{EqContHamil} is then expressed through
finite differences:
\begin{equation}
\frac{\partial^2}{\partial\phi^2}v(\phi)\rightarrow\frac{v_{i-1}-2
v_i+ v_{i+1}}{\delta_{\phi}^2}
\end{equation}
Eventually, the Schr\"odinger equation can be written in a
discrete manner:
\begin{equation}
\frac{v_{i-1}-2v_i+v_{i+1}}{\delta_{\phi}^2}+W_iv_i = \varepsilon
v_i
\end{equation}
Taking into account periodic boundary conditions
($\phi_{n+1}\stackrel{!}{=}\phi_1$), the resulting matrix can be
diagonalized by standard algebraic methods (with eigenvalues
$\varepsilon$).

In the present work, the lowest 30 energies and wavefunctions were
calculated for both the GS and ES torsional potentials of 3P.
Subsequently, the FC overlap integrals have been evaluated by
numerical integration of the respective librational wavefunctions.

\section{\label{Result}Results and Discussion}
\subsection{\label{ExpRes}Experimental spectra}
\begin{figure}[!h]
\begin{center}
\includegraphics[scale=0.7]{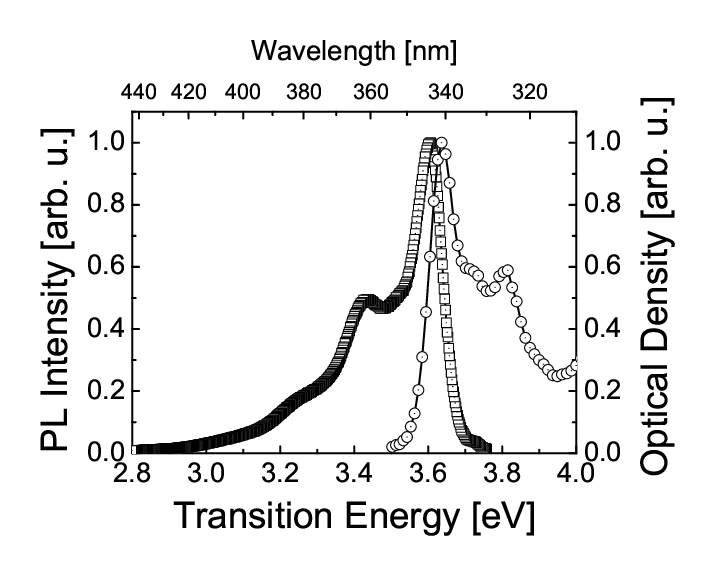}
\caption{\label{FigExp3F} Experimental absorption (circles) and
emission (squares) spectra of indenofluorene in $CH_2Cl_2$
solution at ambient temperature.}
\end{center}
\end{figure}

\begin{figure}[!h]
\begin{center}
\includegraphics[scale=0.7]{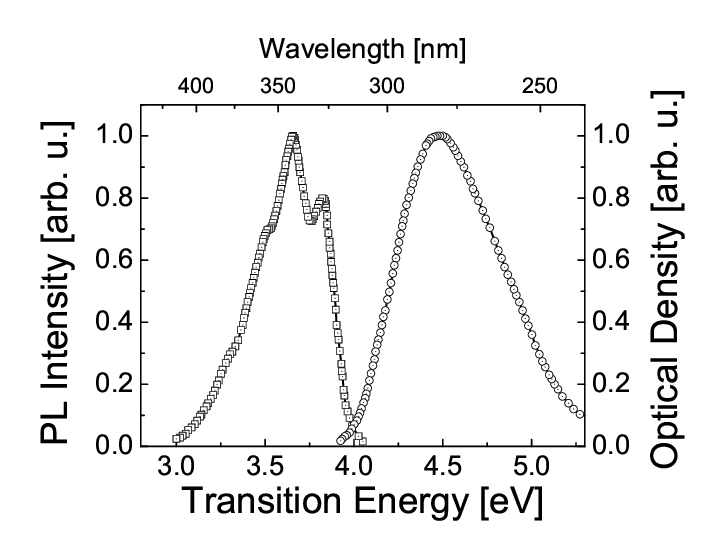}
\caption{\label{FigExp3P} Experimental absorption (circles) and
emission (squares) spectra of \textit{p}-terphenyl in cyclohexane
solution at ambient temperature taken from Ref.
\onlinecite{OppOptics}.}
\end{center}
\end{figure}

The room-temperature experimental absorption and emission spectra
of 3F and 3P are shown in Figs. \ref{FigExp3F} and \ref{FigExp3P}.
For 3F, the mirror symmetry between absorption and emission is
largely conserved. The deviations on the high energy side of the
absorption spectrum are due to superimposed transitions to higher
lying electronic states. Both emission and absorption exhibit a
sharply resolved vibronic structure. In contrast, 3P features an
entirely unstructured, broad absorption band as opposed to a much
better resolved vibronic structure in emission.
Moreover, the Stokes shift between the maxima of absorption and
emission spectra in 3P ($\sim 0.83$ eV) is substantially larger
than in 3F ($\sim 0.04$ eV), suggesting a much larger
reorganization energy in the ES.

\subsection{\label{CalcRes}Calculated bandshapes}
\subsubsection{\label{Calc3Lopp}Indenofluorene}
\begin{figure}[!b]
\begin{center}
\includegraphics[scale=0.7]{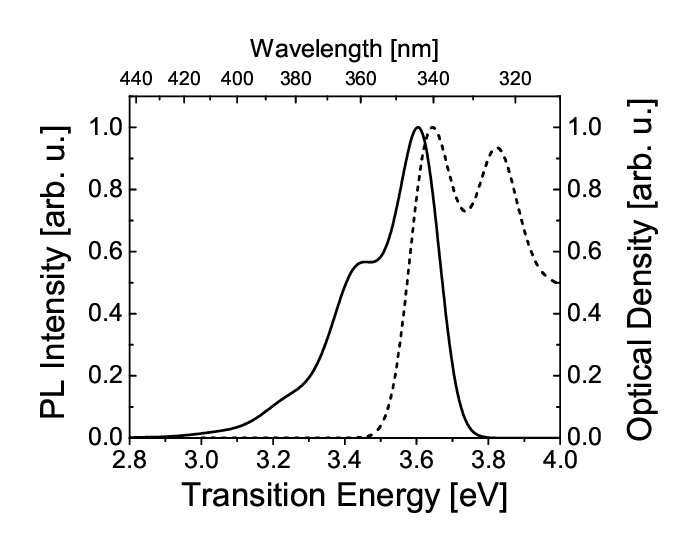}
\caption{\label{FigCalc3F} Calculated spectral bandshapes for
absorption (dashed line) and emission (solid line) of
indenofluorene (3F).}
\end{center}
\end{figure}

Following the procedure outlined in Sections \ref{Theory} and
\ref{Method}, the major five contributing vibrational modes have
been identified for 3F. The respective (unscaled) GS and ES
frequencies and reduced masses are listed in Tab.
\ref{Tab3Fmodes}.
\begin{table}[!ht]
\caption{\label{Tab3Fmodes} Unscaled vibrational frequencies
$\omega_i$ (in cm$^{-1}$), effective masses $\bar{m}_i$ (in atomic
mass units), and mass weighted projections $\Delta Q_i$ of the
ground to excited state distortion (in $\sqrt{amu}$\AA) of the
five most important vibrations for indenofluorene. The respective
Duschinsky matrix elements $D_{i'i''}$ are also listed.}
\begin{center}
\begin{ruledtabular}
\begin{tabular}{crrrcrrrr}
 & \multicolumn{3}{c}{\textbf{Ground State}} & & \multicolumn{3}{c}{\textbf{Excited State}} & \\
\cline{2-4}\cline{6-8}
\raisebox{1.5ex}[-1.5ex]{\textbf{Mode No.}} & $\omega_i'$ & $\bar{m}_i'$ & $\Delta Q_i'$ & & $\omega_i''$ & $\bar{m}_i''$ & $\Delta Q_i''$ & \raisebox{1.5ex}[-1.5ex]{$D_{i'i''}$}\\
\hline
\textit{1} & 221 & 5.49 & 0.43 & & 217 & 5.51 & -0.42 & 0.99\\
\textit{2} & 555 & 4.29 & -0.11 & & 542 & 4.61 & 0.13 & 0.98\\
\textit{3} & 821 & 6.09 & 0.17 & & 784 & 6.11 & -0.18 & 0.99\\
\textit{4} & 1455 & 2.02 & -0.10 & & 1473 & 2.77 & 0.12 & 0.65\\
\textit{5} & 1807 & 6.93 & 0.12 & & 1786 & 6.66 & -0.15 & 0.81\\
\end{tabular}
\end{ruledtabular}
\end{center}
\end{table}
Overall, the normal coordinates of GS and ES are found to coincide
($D_{i'i''}\approx 1$). The GS to ES Duschinsky rotation of mode
no. 4 does not dramatically influence the outcome of the
calculation since it has the smallest GS to ES distortion ($\Delta
Q_4$) associated with it. Inserting these results into Eq.
\ref{EqFcFactors2} allows us to evaluate the FC weights of all
transitions between the vibrational manifolds of the GS and the
ES. The resulting calculated bandshapes (Eq. \ref{EqFullProb}) for
absorption and emission at 295 K are shown in Fig.
\ref{FigCalc3F}.
The electronic origin has been set to the experimental value of
3.62 eV. The Stokes shift between the maxima of the two spectra
(calculated to be $\sim 0.04$ eV) is a direct result of the
calculation. We find good overall agreement with the measured
spectra shown in Fig. \ref{FigExp3F}. Comparison with experiment
indicates that the inhomogeneous line broadening is smaller in 3F
than it is in 3P. Reducing the damping factor in the calculated
spectra of 3F reveals more details in the vibronic structure
(\textit{e.g.}, the shoulders at $\sim 3.5$ eV and $\sim 3.7$ eV
in Fig. \ref{FigExp3F}), all of which are fully consistent with
the experimental results. However, a broadening of 0.1 eV has been
assumed in the calculations in order to stay consistent with the
results presented for 3P. The slight overestimation of the
intensity of the 0-1 transition in absorption can be attributed to
the (partial) neglect of Duschinsky rotation and/or the inaccuracy
($\sim 10 \%$) of CIS force constants.

The most important result is that both absorption and emission
spectra exhibit a sharply resolved vibronic progression. In
accordance with experiment, the two spectra obey mirror image
symmetry to a large extent.

\subsubsection{\label{Plan3p} "Planar" \textit{p}-terphenyl}
Applying the same computational procedure to 3P, one can again
identify the five in-plane high-frequency modes that couple most
strongly to the optical transition. Their respective (unscaled) GS
and ES frequencies and reduced masses are listed in Tab.
\ref{Tab3Pmodes}.
\begin{table}[!h]
\caption{\label{Tab3Pmodes} Unscaled vibrational frequencies
$\omega_i$ (in cm$^{-1}$), effective masses $\bar{m}_i$ (in atomic
mass units), and mass weighted projections $\Delta Q_i$ of the
ground to excited state distortion (in $\sqrt{amu}$\AA) on the
five most important, high-energy, harmonic backbone vibrations for
\textit{p}-terphenyl. The respective Duschinsky matrix elements
$D_{i'i''}$ are also listed.}
\begin{center}
\begin{ruledtabular}
\begin{tabular}{crrrcrrrr}
 & \multicolumn{3}{c}{\textbf{Ground State}} & & \multicolumn{3}{c}{\textbf{Excited State}} & \\
\cline{2-4}\cline{6-8}
\raisebox{1.5ex}[-1.5ex]{\textbf{Mode No.}} & $\omega_i'$ & $\bar{m}_i'$ & $\Delta Q_i'$ & & $\omega_i''$ & $\bar{m}_i''$ & $\Delta Q_i''$ & \raisebox{1.5ex}[-1.5ex]{$D_{i'i''}$}\\
\hline
\textit{1} & 224 & 6.72 & -0.13 & & 243 & 6.93 & 0.13 & 1.00\\
\textit{2} & 838 & 5.99 & 0.04 & & 811 & 6.15 & -0.11 & 0.83\\
\textit{3} & 1301 & 1.14 & 0.16 & & 1359 & 1.16 & -0.20 & 0.90\\
\textit{4} & 1391 & 4.43 & 0.16 & & 1460 & 4.47 & -0.15 & 0.98\\
\textit{5} & 1804 & 5.62 & 0.18 & & 1795 & 5.17 & -0.17 & 0.97\\
\end{tabular}
\end{ruledtabular}
\end{center}
\end{table}

\begin{figure}[!t]
\begin{center}
\includegraphics[scale=0.7]{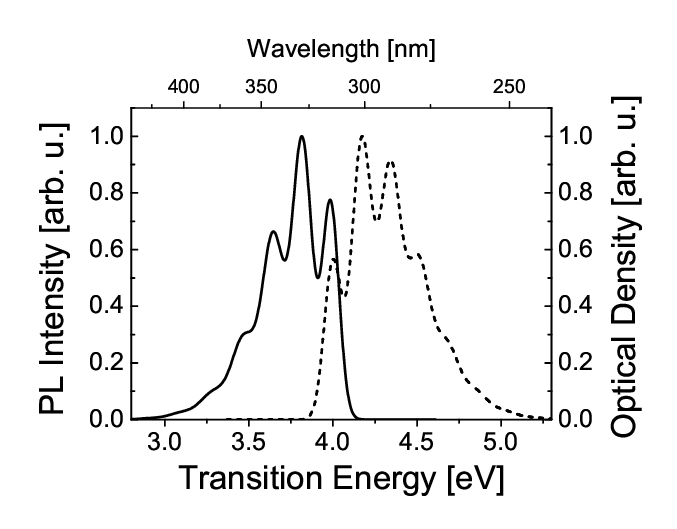}
\caption{\label{FigCalc3Pwo} Calculated spectral bandshapes for
absorption (dashed line) and emission (solid line) in
\textit{p}-terphenyl (3P). The phenyl ring librations are not
taken into account.}
\end{center}
\end{figure}

In this first approach, the ring torsional vibrations are
completely neglected. The respective calculated bandshapes for
absorption and emission are presented in Fig. \ref{FigCalc3Pwo}
for 295 K with the electronic origin set to the experimentally
determined value of 3.99 eV.

The spectra in Fig. \ref{FigCalc3Pwo} nicely obey the mirror image
rule and show a clearly resolved vibronic structure both in
emission and in absorption. They do not, however, reproduce the
experimental findings shown in Fig. \ref{FigExp3P}.

\subsubsection{\label{Harm3P} P-terphenyl in the harmonic
approximation} As opposed to 3F, which features a rigid backbone,
the individual phenylene rings of 3P are free to librate around
the long molecular axis. In the electronic GS, the competition
between steric hindrance of the \textit{ortho}-hydrogens and
conjugation which tends to planarize the molecule leads to a
double-well, \textit{W}-shaped potential for the ring torsional
motion.\cite{DoubleWellRev} Due to changes in the electronic
structure of 3P upon photoexcitation, the inter-ring bonds are
strengthened in the electronic ES, thus overcoming the steric
hindrance imposed by the \textit{ortho}-hydrogens. Overall, this
leads to a planar structure in the ES stabilized by an (almost)
parabolic, \textit{U}-shaped torsional potential. As a consequence
of the substantial change of the inter-ring tilt angle between the
equilibrium conformations in the GS and the ES, the librational
mode strongly couples to the optical transition.
\begin{figure}[!b]
\begin{center}
\includegraphics[scale=0.8]{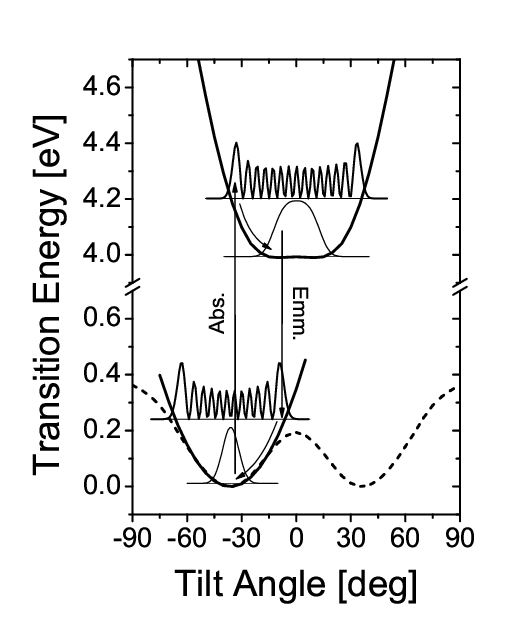}
\caption{\label{FigHarmApprox} The potential energy curves
associated with the libration of the phenyl rings in the
electronic ground (GS) and excited state of \textit{p}-terphenyl
(3P). The true GS potential is represented by the dashed curve,
the harmonic approximation is solid. Probability amplitudes are
plotted for the initial (thin solid line) and most important final
state (thick solid line) for absorption and emission (within
harmonic approximation).}
\end{center}
\end{figure}

In order to demonstrate the importance of a proper consideration
of the actual shape of the torsional potential, we first report
the results obtained when adopting the commonly applied harmonic
approximation for the GS torsional potential well.\cite{OppFc1} To
this end, one branch of the torsional double-well is approximated
by a parabola centered at one of the double-well's minima (see
Fig. \ref{FigHarmApprox}). For reasons of consistency, the ES
potential has been taken as calculated. Subsequently, the
numerically procedure outlined in Subsection \ref{CompMeth} has
been applied in order to evaluate the transition energies and FC
overlap matrix elements.

\begin{figure}[!b]
\begin{center}
\includegraphics[scale=0.8]{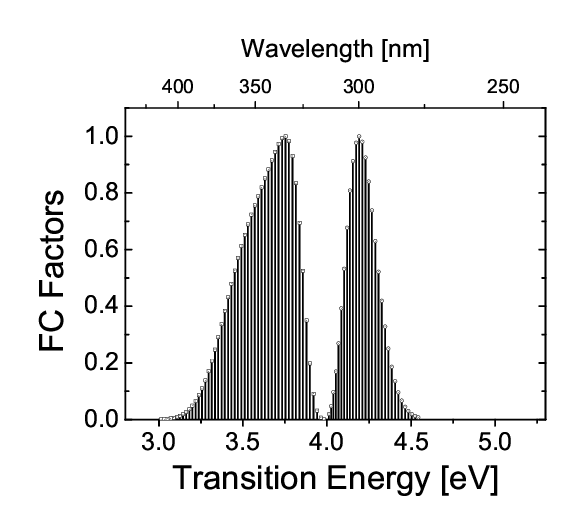}
\caption{\label{FigHarmProg} Vibronic progression associated with
the phenyl ring librations in \textit{p}-terphenyl (3P) within the
harmonic approximation for the torsional potential of the
electronic ground state at 0 K.}
\end{center}
\end{figure}

The situation sketched in Fig. \ref{FigHarmApprox} is (almost)
symmetrical for absorption and emission. Given the large GS to ES
displacement ($\Delta Q=\sqrt{I}\Delta\phi$) associated with the
torsional vibration, one would expect a far extended vibronic
progression of finely separated lines with (almost) Poisson
distributed intensities.\cite{OppFc1,FcCalcs1,Gierschner} In order
to illustrate the contribution to the total bandshapes coming from
the ring torsional vibration within the harmonic approximation to
the GS potential well, the associated vibronic progression is
plotted in Fig. \ref{FigHarmProg} for 0 K.
As indicated in Figs. \ref{FigHarmApprox} and \ref{FigHarmProg},
the most important accepting librational states have high quantum
numbers of $v'=11$ for emission and $v''=14$ for absorption. The
latter value is close to the $\tilde{v}''=11$ reported for
biphenyl in supersonic jet expansion
spectroscopy.\cite{2pTorsionExpSsJet}

The energetic splitting of 100-170 cm$^{-1}$ (see Ref.
\onlinecite{2pTorsionExpSsJet}) between the individual transitions
within the subset of the librational levels is too small to be
resolved experimentally in solution. Because each of the
well-resolved peaks stemming from the 5 high-energy harmonic
backbone modes (see Fig. \ref{FigCalc3Pwo}) is replaced by the
respective librational progression (Fig. \ref{FigHarmProg}), this
results in additional, almost Gaussian broadening. Due to the
higher curvature of the approximate (harmonic) GS torsional
potential (wider spacing of levels) compared to the ES potential,
this broadening affects the emission spectrum even more severely
than the absorption.

\begin{figure}[!t]
\begin{center}
\includegraphics[scale=0.8]{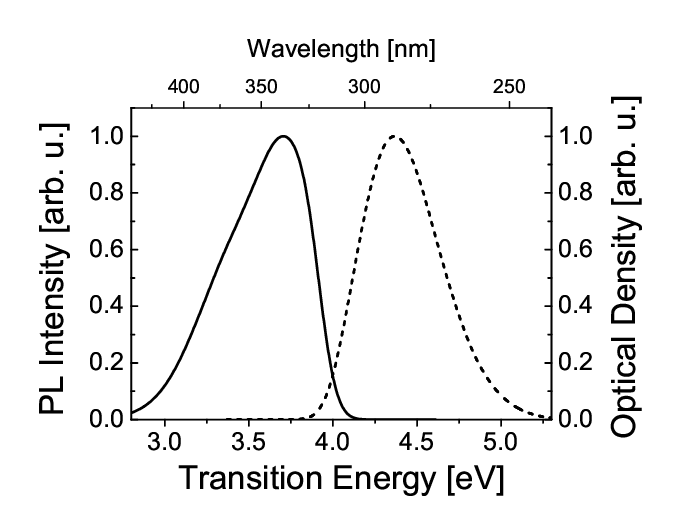}
\caption{\label{FigHarmApproxSpec} Calculated spectral bandshapes
for absorption (dashed line) and emission (solid line) in
\textit{p}-terphenyl (3P). The phenyl ring librations have been
taken into account using the harmonic approximation for the
torsional potential in the electronic ground state.}
\end{center}
\end{figure}

The calculated bandshapes that take into account both the 5
high-energy modes (Tab. \ref{Tab3Pmodes}) and the inter-ring
libration in its harmonic approximation are shown in \mbox{Fig.
\ref{FigHarmApproxSpec}}.
In this case, both the emission and absorption spectra are equally
smeared out to featureless bands, exhibiting a large Stokes shift,
as has been predicted in Ref. \onlinecite{FcCalcs1}. They do,
however, obey mirror image symmetry which is in clear
contradiction to the experimental observations (Fig.
\ref{FigExp3P}).

\subsubsection{Anharmonic torsion in p-terphenyl}
\begin{figure}[!t]
\begin{center}
\includegraphics[scale=0.8]{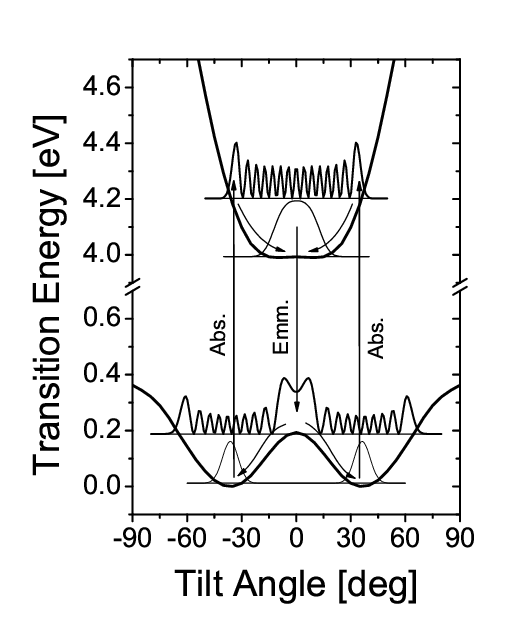}
\caption{\label{FigTruePot} The actual potential energy curves
associated with the libration of the phenyl rings in the
electronic ground (GS) and excited state of \textit{p}-terphenyl
(3P). Probability amplitudes are plotted for the initial (thin
solid line) and most important final state (thick solid line) for
absorption and emission.}
\end{center}
\end{figure}

Only the inclusion of the actual, strongly anharmonic double-well
potential for the GS torsion can account for the breakdown of the
mirror image symmetry in oligo(\textit{para}-phenylene)s in
general and in 3P in particular, as pointed out in Ref.
\onlinecite{OppFc2}. The massive deviation of the GS potential
from the harmonic approximation introduces the necessary asymmetry
between the accepting librational states in GS and ES. The nearly
harmonic ES potential features harmonic oscillator-like
vibrational wavefunctions (see Figs. \ref{FigTruePot} and
\ref{FigHarmApprox}), which results in a wide vibronic progression
associated with the ring torsion that smears out the
\emph{absorption} spectrum to an unstructured, broad band. In
contrast, the strongly anharmonic potential in the GS leads to
wavefunctions that deviate significantly from harmonic oscillator
wavefunctions (compare Figs. \ref{FigTruePot} and
\ref{FigHarmApprox}).
This leads to a massive deviation of the associated vibronic
progression from the Poisson-type envelop for \emph{emission}.

\begin{figure}[!t]
\begin{center}
\includegraphics[scale=0.8]{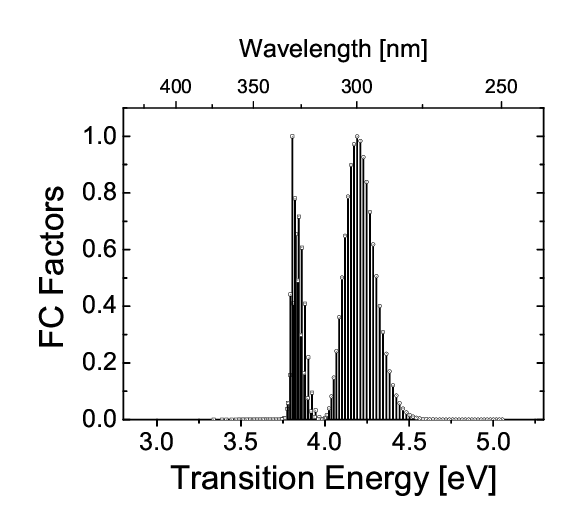}
\caption{\label{FigTorsProg} Vibronic progression associated with
the phenyl ring librations in \textit{p}-terphenyl (3P) for the
actual, librational double-well potential in the electronic ground
state at 0 K.}
\end{center}
\end{figure}

To illustrate this feature, the progressions related to the phenyl
ring librations are plotted in Fig. \ref{FigTorsProg} for
absorption and emission at 0 K, taking full account of the actual
GS torsional potential.
The librational progression for \emph{absorption} from the actual
GS potential equals that from the harmonic approximation. The
dominant accepting state is also that with $v''=14$. For
\emph{emission}, the most important accepting libronic state in
the GS has a quantum number of $v'=9$ and energetically lies just
above the central potential barrier, thus exhibiting a probability
maximum at $0^\circ$. Despite its high quantum number, the
functional form of the associated wavefunction is somehow
reminiscent of that of a harmonic oscillator ground-state function
for a (harmonic) potential centered around 0$^{\circ}$. This
explains also why the overlap with the $v''=0$ librational
wavefunction in the ES (also shown in Fig. \ref{FigTruePot}) is
largest. Effectively, the situation for \emph{emission} resembles
a system in which the libration does not couple at all to the
optical transition (i.e., one in which the GS to ES interstate
distortion along the associated normal coordinate is zero). This
is reflected by the very narrow libronic progression for
\emph{emission} centered around the dominant $v''=0\rightarrow
v'=9$ peak.

To describe the optical spectra, again each of the well-resolved
peaks stemming from the 5 high-energy harmonic backbone modes (see
Fig. \ref{FigCalc3Pwo}) is replaced by the respective librational
progression (Fig. \ref{FigTorsProg}). Taking the actual GS
torsional potential into account, the sharp transitions get
smeared out in \emph{absorption} a lot more than in
\emph{emission}. This is a direct consequence of the combination
of the (anharmonic) GS double-well with the (harmonic) ES single
well, regardless of their exact shape.

\begin{figure}[!b]
\begin{center}
\includegraphics[scale=0.8]{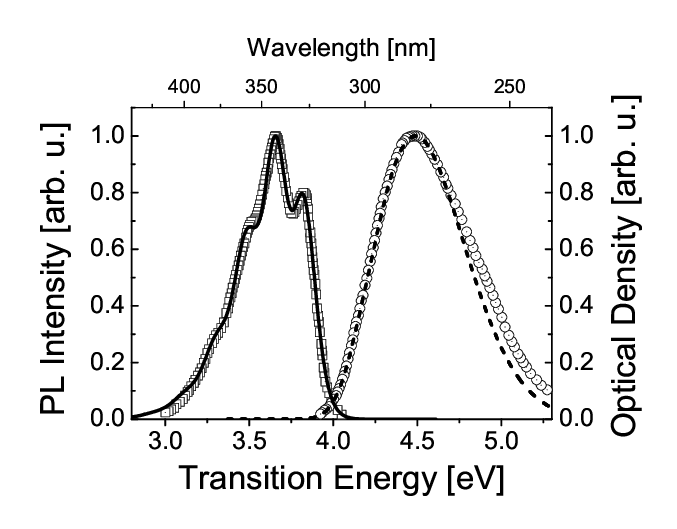}
\caption{\label{FigFinalSpec} Calculated absorption (dashed line)
and emission (solid line) bandshapes for \textit{p}-terphenyl
taking into account the anharmonic double-well potential for the
phenyl ring librations. The experimental absorption and emission
spectra are represented by circles and squares, respectively.}
\end{center}
\end{figure}
Combining the results shown in Fig.\ref{FigTorsProg} with those
presented in Fig. \ref{FigCalc3Pwo} yields the total bandshapes
for absorption and emission in 3P. These are plotted in Fig.
\ref{FigFinalSpec} together with the experimental data from Fig.
\ref{FigExp3P}. Here again, only the electronic origin has been
shifted to the experimentally determined value of 3.99 eV, whereas
the Stokes shift (calculated to be $\sim 0.83$ eV) is a direct
result of our calculations.

Comparing the optical spectra of 3F and 3P shows that the Stokes
shift is substantially larger in the latter. This implies that the
reorganization energy in the excited state and thus the additional
stabilization of the exciton is strongly increased by the presence
of strongly coupled low-energy conformational degrees of freedom.
This is fully consistent with the results of our calculations,
which yield a reorganization energy of 0.34 eV in 3F and 2.5 times
as much, 0.84 eV, for 3P.

\section{\label{Outlook}Related systems and low-temperature optical spectra}
In solution, not only 3P but also all other
oligo(\textit{para}-phenylene)s (from biphenyl to
\textit{p}-sexiphenyl) exhibit the same broad, featureless
absorption band as opposed to a vibronically well-structured
emission.\cite{OppOptics} Apart from the fact that starting with
\textit{p}-quaterphenyl there are more than one symmetric ring
torsional vibration present in the system, we suggest that the
mechanism outlined in the present work can account for the
observed breakdown of the mirror image symmetry in all
oligo(\textit{para}-phenylene)s.

\begin{figure}[!b]
\begin{center}
\includegraphics[scale=0.8]{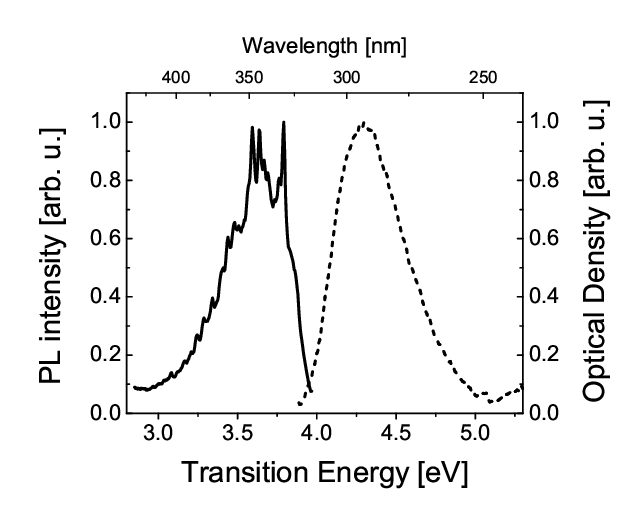}
\caption{\label{FigExpLtSpec3p} Experimental absorption (dashed
line) and emission (solid line) spectra of \textit{p}-terphenyl in
\textit{n}-decane solution recorded at 15 K.}
\end{center}
\end{figure}

In contrast to PPV oligomers (\textit{vide infra}), the observed
asymmetry between the absorption and emission spectra in 3P is
conserved also at low temperatures (see Fig.
\ref{FigExpLtSpec3p}).
This further supports our conclusion that the breakdown of the
mirror image symmetry in 3P is indeed an \emph{intrinsic} effect
that is only related to the particular combination of the GS
(\textit{W}-shaped) and ES (\textit{U}-shaped) torsional
potentials and not to different thermal broadening of absorption
and emission (we have not included calculated spectra for low
temperature since the torsional potential wells used to calculate
the spectral bandshapes in solution cannot be expected to
correctly describe the rather complex situation\cite{Matrix1} at
low temperatures in solidified solution/matrix).

As mentioned in the introduction, PPP\cite{Reli1} and
poly(fluorene)\cite{PfoBands1,PfoBands2,PfoBands3,PfoBands4,EmilPfoReview}
also exhibit one broad unstructured absorption band and a
well-resolved vibronic progression in emission, both in solution
and in the solid state. Given the similarity of the chemical
structures (phenylene rings \textit{para}-linked by a C-C single
bond), the double-well potential in the GS for the ring libration
can be expected to prevail here. The same reasoning as for 3P can
then be applied to these polymeric systems. Of course, effects
related to excitation energy migration also play a role in
polymeric samples with a broad distribution of effective
conjugation lengths, as discussed in the introduction.

Temperature dependent optical studies on thin films of
poly(9,9-(di-\textit{n},\textit{n}-octyl)fluorene) revealed that
in the so-called $\beta$-phase, the side chains crystallize,
rigidly confining the polymer backbone to a planar
conformation.\cite{Matrix6,PfoBeta1,PfoBeta2,PfoBeta3,EmilPfoReview}
Librations of the individual repeat units are thus effectively
hindered. As a consequence, the absorption spectrum associated
with the $\beta$-phase exhibits a well-pronounced vibronic
structure associated with high-energy, in-plane C=C stretch modes,
similar to that observed in
emission.\cite{PfoBands2,PfoBands3,PfoBands4,EmilPfoReview}

A number of temperature dependent optical measurements have also
been performed on (unsubstituted) PPV oligomers in (frozen)
solution and
matrix.\cite{Gierschner,Matrix2,Matrix3,Matrix4,Matrix5,Matrix7}
The overall finding was that broadening of the backbone mode
progressions significantly decreases upon decreasing temperature.
The absorption spectrum evolves from a broad, relatively
featureless band towards the mirror image of the sharply
structured emission. The strong broadening of the absorption as
compared to emission at ambient temperature has been attributed to
thermal population of the librational levels (energy spacing wider
in the ES than in the GS)\cite{Gierschner}. The PPV oligomers can
be distinguished from the oligo(\textit{para}-phenylene)s in that
the former show a \emph{planar} equilibrium geometry in the GS
together with a high torsional flexibility,\cite{Gierschner}
whereas the latter exhibit a \emph{non-planar} GS equilibrium
geometry. Due to this qualitatively different form of the
torsional potential wells in PPV oligomers (both \emph{harmonic}
single-wells centered at $0^{\circ}$, only different curvature in
GS and ES), the slightly more involved approach pursued in this
work is not necessary (though applicable) to satisfactorily model
their optical spectra (compare Ref. \onlinecite{Gierschner}).
Experimentally, the difference between these two types of
molecules (PPV and oligo(\textit{para}-phenylene)s) cannot be told
from the room-temperature optical spectra; it shows very clearly,
however, at low temperature.

\section{\label{Conclusion}Conclusion}
By comparing the \emph{soft} backbone \textit{p}-terphenyl with
the rigidly planar but otherwise identical indenofluorene, the
violation of the mirror image symmetry in the optical spectra of
\textit{p}-terphenyl has been investigated. Applying
\textit{ab-initio} quantum chemical methods and further numerical
treatment, the vibronically resolved bandshapes of the absorption
and emission spectra of the investigated compounds have been
calculated. The contribution of the phenylene ring librations in
3P has been simulated by numerical evaluation of the related
wavefuntions and Franck-Condon factors in internal curvilinear
coordinates, taking fully into account the substantial
anharmonicity of the related potential wells.

Good agreement with experiment was achieved. A clear correlation
between the torsional degree of freedom and the breakdown of
mirror image symmetry could be established. It has been shown that
only a rigorous Franck-Condon treatment that fully takes into
account the strongly anharmonic double-well potential for the
phenyl ring librations in \textit{p}-terphenyl reproduces the
experimental results. Furthermore, the failure of the harmonic
approximation for shallow, massively anharmonic potential wells
has been addressed.

Temperature dependent optical measurements on related compounds
have been discussed following the lines of thought of the present
work.

\begin{acknowledgements} The authors would like
to acknowledge the financial support of the SFB Elektroaktive
Stoffe and project P14237-PHY of the Austrian Fonds zur
F\"orderung der Wissnschaftlichen Forschung (FWF). The work at
Georgia Tech is partly supported by the US National Science
Foundation (CHE-034342). The work in Mons and in T\"ubingen was
supported by the European Commission through the Human Potential
Programme (RTN 'Nanochannel', Contract No. HPRN-CT-2002-00323).
The work in Mons was also partly supported by the Belgian National
Fund for Scientific Research (FNRS/FRFC) and by the Belgian
Federal Government in the framework of the ``P\^ole d'attraction
Interuniversitaire en Chimie Supramol\'eculaire et Catalyse
Supramol\'eculaire (PAI5/3)''. CDL-AFM is an important part of the
long term AT\&S research activities.
\end{acknowledgements}

\bibliographystyle{apsrev}


\printtables

\printfigures

\end{document}